\documentclass{emulateapj}

\shorttitle{Evidence of a Plasmoid-Looptop Interaction and Magnetic Inflows During a Solar Flare/CME Eruptive Event}
\shortauthors{Milligan, McAteer, Dennis, \& Young}

\begin{document}

\title{Evidence of a Plasmoid-Looptop Interaction and Magnetic Inflows During a Solar Flare/CME Eruptive Event}

\author{Ryan O. Milligan\altaffilmark{1}, R. T. James McAteer\altaffilmark{2}, Brian R. Dennis\altaffilmark{1} \& C. Alex Young\altaffilmark{3}}

\altaffiltext{1}{Solar Physics Laboratory (Code 671), Heliophysics Science Division, NASA Goddard Space Flight Center, Greenbelt, MD 20771, USA}
\altaffiltext{2}{School of Physics, Trinity College Dublin, College Green, Dublin 2, Ireland}
\altaffiltext{3}{ADNET Systems, Inc., NASA Goddard Space Flight Center, (Code 671), Greenbelt, MD 20771, USA}

\begin{abstract}
Observational evidence is presented for the merging of a downward-propagating plasmoid with a looptop kernel during an occulted limb event on 2007 January 25. RHESSI lightcurves in the 9--18~keV energy range, as well as that of the 245 MHz channel of the Learmonth Solar Observatory, show enhanced nonthermal emission in the corona at the time of the merging suggesting that additional particle acceleration took place. This was attributed to a secondary episode of reconnection in the current sheet that formed between the two merging sources. RHESSI images were used to establish a mean downward velocity of the plasmoid of 12~km~s$^{-1}$. Complementary observations from the SECCHI suite of instruments onboard STEREO-Behind showed that this process occurred during the acceleration phase of the associated CME. From wavelet-enhanced EUVI, images evidence of inflowing magnetic field lines prior to the CME eruption is also presented. The derived inflow velocity was found to be 1.5~km~s$^{-1}$. This combination of observations supports a recent numerical simulation of plasmoid formation, propagation and subsequent particle acceleration due to the tearing mode instability during current sheet formation.
\end{abstract}


\clearpage
\newpage

\section{INTRODUCTION} 
\label{intro} 


During an eruptive event comprising a solar flare and a coronal mass ejection (CME), energy is believed to be converted into the heating of plasma and the kinetic energy of particles and the CME itself through the process of magnetic reconnection. The standard reconnection models (\citealt{park57,swee58,pets64}) state that newly connected field lines expel plasma from the reconnection site due to the Lorentz force. The pressure gradient across the diffusion region then forces new plasma inwards, along with the field lines frozen to it where they change connectivity and dissipate energy. \citet{lin00} stated that these inflows are concurrent with the eruption of a CME, which remains connected to the magnetic neutral point by an extended current sheet. Initially the CME rises slowly until no neighboring equilibrium state is available. After reaching this point the CME begins to rise at an increasing rate. Energy release and particle acceleration continue due to sustained reconnection as the CME accelerates. 

To date, there has been little observational evidence for the predicted inflows associated with reconnection. The most cited example is that of \cite{yoko01}, who found inflow velocities of 1.0--4.7~km~s$^{-1}$ by tracing the movement of threadlike patterns above the limb in a series of {\it SOHO}/EIT 195\AA~images (\ion{Fe}{12}; \citealt{dela95}). Evidence for sustained energy release during CME acceleration has been reported in a recent study of two fast ($>$1000~km~s$^{-1}$) halo CMEs by \cite{temm08,temm10}, who found a strong correlation between the CME acceleration and flare hard X-ray (HXR) time profiles. The bremsstrahlung hard X-rays are signatures of thick-target collisions between the accelerated electrons and the ambient chromospheric material. The authors interpret this correlation as strong evidence for a feedback relationship occurring between CME acceleration and the energy released by magnetic reconnection in the current sheet formed behind the CME. In cases where the current sheet becomes sufficiently longer than its width, it is possible for multiple X-points to form due to the tearing mode instability which can result in the formation of plasmoids \citep{furt63}. Both upward- \citep{ko03,lin05} and downward-directed \citep{shee02} plasmoids associated with CME eruption have been commonly observed in white light coronagraph images, in agreement with MHD simulations (e.g., \citealt{forb83}). Further evidence for plasmoid motions has been presented through radio observations of drifting pulsating structures \citep[DPS;][]{klie00,karl04,rile07,bart08b}.

\begin{figure}[!t]
\begin{center}
\includegraphics[height=8.5cm,angle=90]{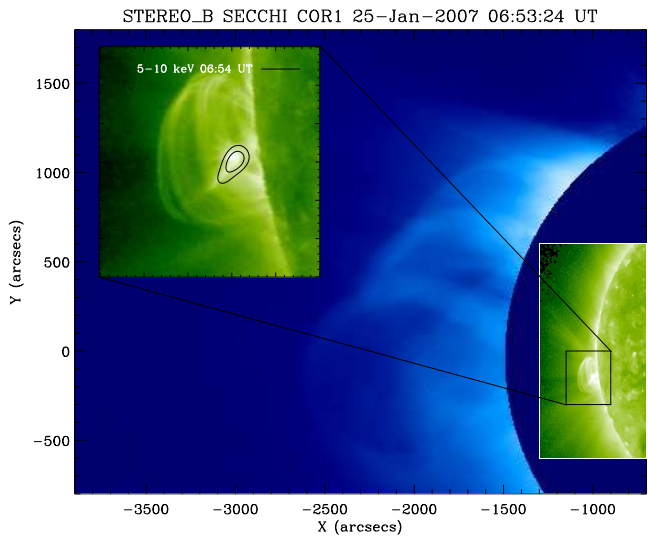}
\caption{The CME on 2007 January 25 as seen by the COR1 coronagraph (blue) at 06:53:24~UT as well as the associated EUVI field of view (green). The expanded box shows the coronal loops that form part of active region NOAA 10940 with 40 and 80\% contours of the 5-10~keV emission seen by RHESSI at the same time (solid line).}
\label{euvi_cor1}
\end{center}
\end{figure}

Observational X-ray evidence for the formation of a current sheet has been presented by \cite{sui03} using data from the Ramaty High Energy Solar Spectroscopic Imager (RHESSI; \citealt{lin02}). The authors were able to show that an above-the-looptop coronal X-ray source (or plasmoid) increased in altitude as a lower lying X-ray loop decreased in altitude during the initial stages of a solar flare. They concluded that magnetic reconnection occurred between the two sources as the current sheet formed. This interpretation was strengthened by evidence that the mean photon energy decreased with distance in both directions away from the reconnection site (see also \citealt{liu08}). The authors attribute the downward moving looptop to the collapse of the X-point to a relaxed magnetic loop during the reconfiguration of the magnetic field. The same conclusions were reached by \cite{sui04} and \cite{vero06}, who observed similar motions of rising plasmoids concurrent with shrinking looptop sources in other events imaged with RHESSI. 

A recent numerical simulation by \cite{bart08a} shows that by invoking variable reconnection rates along the current sheet, {\it downward} propagating plasmoids should also be visible in X-rays below $\sim$2~R$_{\odot}$ (see also \citealt{rile07}). The condition for this scenario is met when the reconnection rate above the plasmoid is greater than that below resulting in a net downward tension in the newly connected magnetic field lines. Furthermore, this model shows that the interaction of such a plasmoid with the underlying loop system can result in a substantial increase in dissipated energy, more so than during the initial ejection of the rising plasmoid or coalescing plasmoid pairs. To date, there has only been one report of such an interaction by \cite{kolo07} using Yohkoh/SXT data. They found an increase in X-ray and decimetric radio flux and an increase in temperature at the interaction site.

\begin{figure}[!t]
\begin{center}
\includegraphics[height=8.5cm,angle=90]{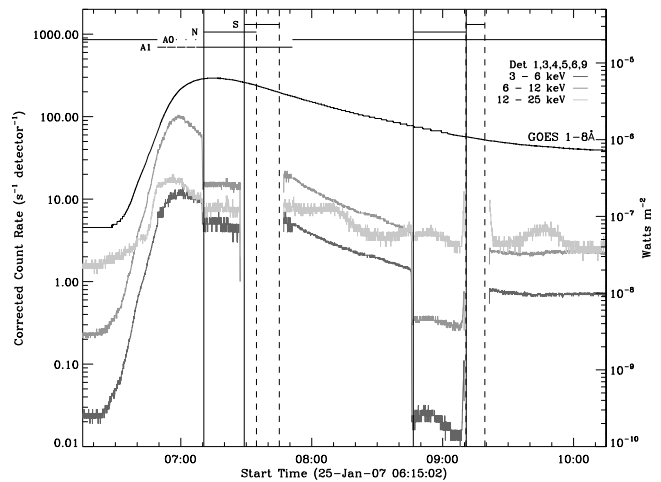}
\caption{Lightcurves of the flare in the 3--6, 6--12, and 12--25~keV energy bands as observed by RHESSI, as well as the GOES 1--8~\AA~lightcurve. The horizontal bars at the top of the plot denote RHESSI's attenuator state (A0, A1), nighttime (N) and SAA passes (S).} 
\label{hsi_goes_ltc}
\end{center}
\end{figure}

\begin{figure*}[]
\begin{center}
\includegraphics[height=\textwidth,angle=90]{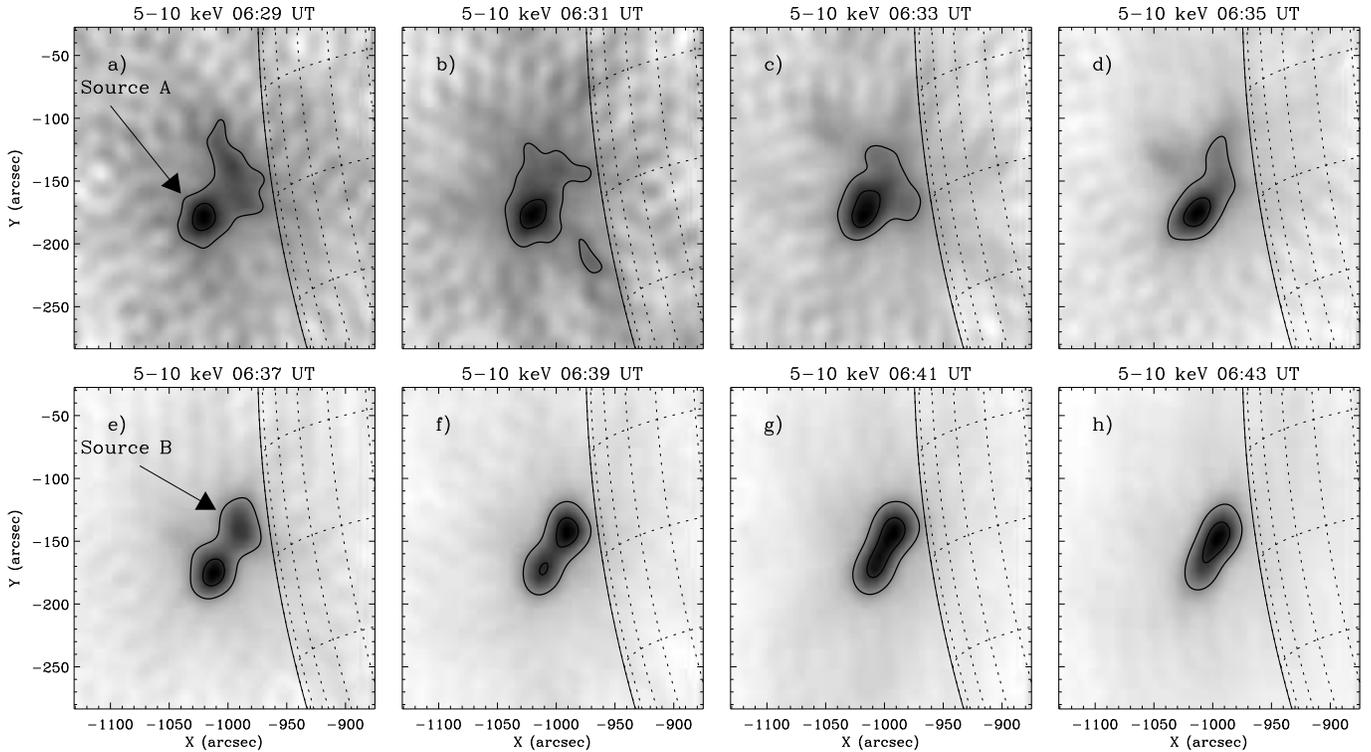}
\caption{RHESSI images in the 5-10 keV energy band formed over 60s integrations during the onset of the flare, although only alternate images are shown here. Contours mark the 40\% and 80\% levels. The plasmoid (source A) and looptop (source B) sources are labeled. The grey pattern around the sources are the CLEAN residuals and reflect the background noise level of the images.}
\label{multi_hsi_plot}
\end{center}
\end{figure*}

\begin{figure}[!b]
\begin{center}
\includegraphics[height=8.5cm]{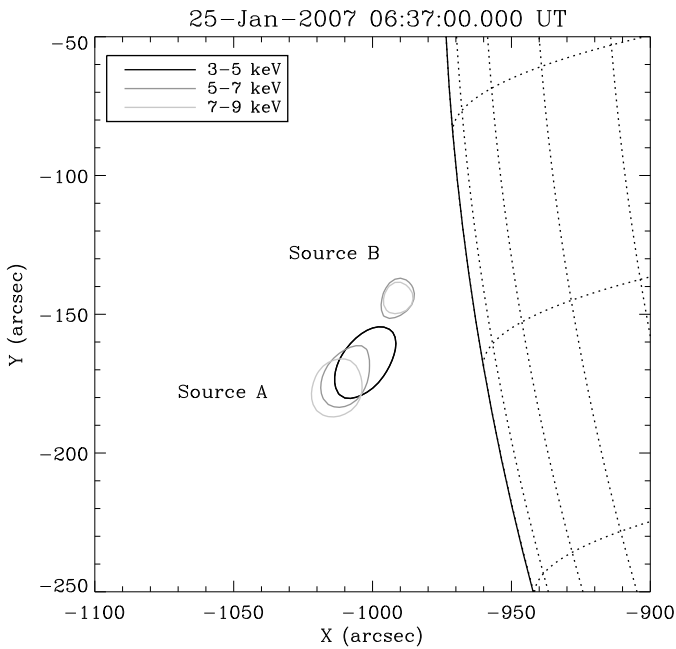}
\caption{The two sources observed by RHESSI imaged over 2~keV wide energy bins (3--5, 5--7, 7--9~keV) for a single time interval.}
\label{hsi_ht_vs_en}
\end{center}
\end{figure}

In this paper observational evidence is presented for increased hard X-ray and radio emission during the coalescence of a downward-moving coronal source with a looptop kernel at the onset of a flare observed with RHESSI. Coordinated observations from the Sun-Earth Connection Coronal and Heliospheric Investigation (SECCHI; \citealt{howa08}) suite of instruments onboard the Solar Terrestrial Earth Relations Observatory (STEREO; \citealt{kais08}) show that this interaction was concurrent with the acceleration phase of the associated CME. Using wavelet enhanced images from the EUV Imager (EUVI), evidence is also presented for inflowing magnetic field lines that persisted for several hours prior to reconnection. Section~\ref{obs} describes the event as viewed by RHESSI and STEREO and the techniques used to determine the motion of the coronal X-ray sources and the CME. Section~\ref{conc} discusses the findings in the context of numerical simulations, and summarizes the conclusions.

\section{OBSERVATIONS AND ANALYSIS}
\label{obs}

The event presented here occurred on 2007 January 25 in active region NOAA 10940, which was just behind the east limb at the time as seen from the Earth. Several CMEs from the same active region were observed around this period shortly after the launch of STEREO, and have been the focus of many studies \citep{attr07,luga08,luga09,grig09}. As STEREO-Behind was only within 0.2$^{\circ}$ from the Sun-Earth line at this time, no corrections were required to align the images with those from RHESSI. Figure~\ref{euvi_cor1} shows the CME as it passed through the COR1 field-of-view at 06:53:24~UT along with the associated active region as seen by EUVI. Also overlaid on the inset EUVI image are the 5--10~keV source contours observed with RHESSI at the same time. Figure~\ref{hsi_goes_ltc} shows the X-ray lightcurves in the 3--6, 6--12, and 12--25~keV energy bands from RHESSI, along with the 1--8~\AA~lightcurve from GOES. The GOES C6.3 class flare began at 06:33:00~UT and peaked at 07:15:00~UT. Emission in the 3--6 and 6--12 keV energy bands observed by RHESSI began to increase $\sim$5 minutes earlier. At the time of this event, there was another active region on the western limb that was the focus of instruments not capable of observing the full solar disk, such as TRACE and those onboard Hinode. Data for the event presented here were, therefore, only obtainable from those instruments capable of observing the entire solar disk. This included radio data from the Learmonth Solar Observatory in Western Australia at eight discreet frequencies  (245, 410, 610, 1415, 2695, 4995, 8800, and 15400 MHz).

\subsection{Coronal X-Ray Source Motions}
\label{x_ray_sources}

RHESSI images were formed using CLEAN \citep{hurf02} in the 5-10~keV energy band over 60s integrations using detectors 4, 6, 8, and 9. Detectors \#2 and 7 were omitted from this analysis due to their reduced sensitivity to low-energy photons. The calibration for detector \#5 was poorly known at this time and was, therefore, also excluded. Detectors \#1, and 3 also introduced noise in the images in this case by over-resolving the sources due to their higher spatial resolution and were therefore also omitted. The 5--10~keV range was chosen to give the best signal to noise ratio below the instrumental background Ge line at $\sim$11~keV during the onset of the flare when the count rate was low. The earliest images revealed a single, high-altitude coronal source (source A; Figure~\ref{multi_hsi_plot}$a$--$f$). At 06:37~UT a second source (source B; Figure~\ref{multi_hsi_plot}$e$--$h$) appeared apparently at a lower altitude than the initial source. Source B was observed to lie above the post-flare loop arcade that later rose from behind the limb in EUVI images (see Figures 3 and 4 in \citealt{grig09}), and was therefore assumed to be a looptop kernel associated with newly formed loops that lay above those emitting in EUV. Source A, on the other hand, resembled an above-the-looptop source or plasmoid. From the bottom row of Figure~\ref{multi_hsi_plot} it can be seen that these two sources merged together between 06:37:00--06:41:00~UT. After 06:41:00~UT the amalgamated source displayed a cusp-like structure extending to the southeast until RHESSI went into the Earth's shadow at 07:09~UT.

Figure~\ref{hsi_ht_vs_en} shows the structure of the two sources in finer energy bins at a time when each of the sources could be clearly resolved (06:37~UT). It is shown that source A displayed an energy gradient in terms of height, with higher energy emission emanating from higher altitudes. Source B on the other hand showed no discernible displacement in terms of energy. This is in contrast to what was observed in the event presented by \cite{kolo07}, who found a thermal stratification in the looptop source, but no clear displacement for the higher altitude source. Similarly, \cite{sui03} found that higher energy emission emanated from higher altitudes for their looptop source, as expected. However, the reverse was found to be true for the associated rising plasmoid with mean photon energy decreasing with height, consistent with the idea that reconnection occurred in between the two sources. Similar conclusions were reached by \cite{liu08} who stated that higher energy emission should be observed closer to the reconnection site \citep{liu08}. With this in mind the case presented in Figure~\ref{hsi_ht_vs_en} suggests that in forming the plasmoid, the reconnection rate above the source must have been greater than that below. This would not only explain the reverse energy stratification as a function height, but also the resulting downward motion due to the resulting net tension exerted by the magnetic field, as surmised by \cite{bart08a}.

In order to track the motion of each source, the coordinates of the peak emission were identified and used to triangulate their height above the solar limb. The peak emission, rather than the centroid, was chosen to exclude the possibility of interpreting the relative change in intensity of the two sources as a motion. It was found that source A had an initial height of 45~Mm at 06:29~UT and decreased in altitude during the subsequent 12 minutes (Figure~\ref{hsi_ltc_ht_radio}$c$). A linear least-squares fit to these data points yielded a mean downward velocity of 12~km~s$^{-1}$, similar to the value of 16~km~s$^{-1}$ found by \citet{kolo07} for their downward-moving plasmoid. Source B was observed to rise continuously throughout the event, which is characteristic of a post-flare arcade. Its mean velocity was found to be $\sim$5~km~s$^{-1}$. After 06:41:00~UT the individual sources could no longer be resolved therefore the time interval over which the two sources merged was estimated to be from 06:37 to 06:41~UT.

\subsection{Evidence for Enhanced Nonthermal Emission}
\label{xray_spec}

\begin{figure}[!t]
\begin{center}
\includegraphics[width=8.5cm]{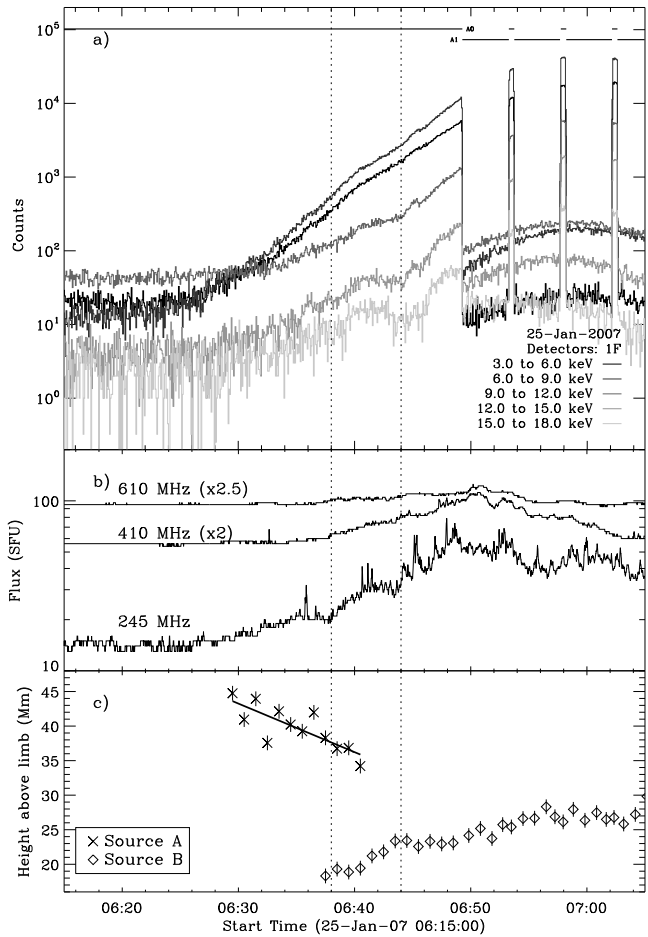}
\caption{$a$) RHESSI lightcurves in the 3--6, 6--9, 9--12, 12--15, and 15--18~keV energy bands from the front segment of detector \#1 only. The horizontal bars marked A0 and A1 at the top of the plot denote the attenuator state. $b$) Emission in the 245, 410 and 610~MHz channels from Learmonth radio telescope. The fluxes in the 410 and 610 MHz channels have been scaled by factors of 2 and 2.5 for clarity, respectively. $c$) Height-time plots of the two 5--10~keV sources as observed by RHESSI. The plasmoid source is denoted by crosses while the looptop source is given by diamonds, both with error bars. The solid line represents a least-sqaures fit to the downward moving source. The two vertical dotted lines mark the approximate time of enhanced HXR and radio emission during which the two RHESSI sources appeared to merge.}
\label{hsi_ltc_ht_radio}
\end{center}
\end{figure}

According to \cite{bart08b}, a plasmoid-looptop interaction as described in Section~\ref{x_ray_sources} should have a distinct observational signature. The authors predict that the resulting increase in energy dissipation should manifest itself as enhanced chromospheric or HXR emission. In the event presented by \cite{kolo07}, the authors observed an concurrent increase in both HXRs (14--23~keV) and radio emission (1--2~GHz), both indicators of nonthermal electrons. The authors also detected an increase in temperature at the interaction site in the corona during the merging. Figure~\ref{hsi_ltc_ht_radio}$a$ shows the RHESSI lightcurves (in raw counts) in 3~keV wide energy bins (3--6, 6--9, 9--12, 12--15, and 15--18~keV) over the flare onset using the front segment of detector \#1 only. Between 06:38 and 06:44~UT (shown by the two vertical dotted lines in Figure~\ref{hsi_ltc_ht_radio}) there is a pronounced enhancement in the higher energy bands (12--15 and 15--18~keV, in particular). A similar enhancement is also visible in the 245 MHz channel of the Learmonth radio data (Figure~\ref{hsi_ltc_ht_radio}b). The increase in emission (from 06:38--06:41~UT) corresponds to the approximate time over which the two X-ray sources were observed to merge from Figure~\ref{multi_hsi_plot}e--g. From 06:41--06:44~UT (after the plasmoid source was no longer visible) HXR and radio emissions both appeared to decrease briefly. This episode of increased nonthermal emission is therefore believed to be a result of a secondary phase of particle acceleration due to magnetic reconnection within the current sheet formed between the two merging sources. Unfortunately there was no radio spectrograph data available at the time of this event to search for evidence of drifting pulsating structures. 

\begin{figure}[!t]
\begin{center}
\includegraphics[width=8.5cm]{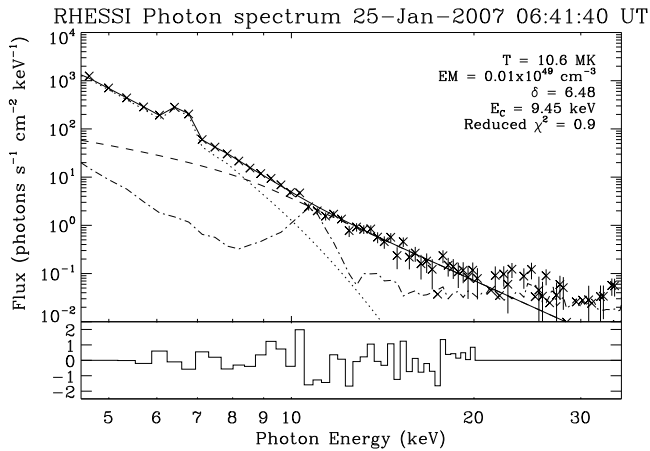}
\caption{RHESSI photon spectra with the associated residuals using the front segment of detector \#1 integrated over 06:41:40--06:42:00~UT during the merging phase. The dotted line represents the best fit to the thermal distribution while the dashed line represents the thick-target component. The solid line shows the sum of the two components and the dot-dashed line marks the background.}
\label{spec_fits}
\end{center}
\end{figure}

A RHESSI spectrum taken around the time of the merging (06:41:00~UT; Figure~\ref{spec_fits}) also shows that emission from 9--20~keV is predominantly nonthermal, consistent with the idea that enhancements in both the HXR and radio lightcurves are evidence for an increase in the number of accelerated particles. This spectrum was also generate using only detector \#1 to remain consistent with the lightcurve shown in Figure~\ref{hsi_ltc_ht_radio}. This increased nonthermal emission is consistent with the simulations of \cite{bart08b} but is clearly coronal in nature, rather than chromospheric as predicted. Chromospheric rebrightening cannot be ruled out however but may be difficult to simultaneously detect both coronal plasmoids and footpoint emission during on-disk events due to RHESSI's limited dynamic range.

\subsection{CME Kinematics}
\label{cme_acc}

One limitation of many previous studies of CMEs is the absence of data below $\sim$3~R$_{\odot}$, where most of the CME acceleration takes place. This is due in part to the loss of the C1 coronagraph on SOHO/LASCO in 1998. With the launch of STEREO in 2006, this gap has been filled with the SECCHI suite of instruments. EUVI captures full-disk observations out to 1.7~R$_{\odot}$, while the COR1 and COR2 coronagraphs cover 1.4--4~R$_{\odot}$ and 2--15~R$_{\odot}$, respectively. 

The data used in this study are exclusively from the STEREO-Behind (STEREO-B) spacecraft and were prepped using the standard {\sc secchi\_prep} routine inside SSWIDL. For EUVI, this entails standard corrections for de-bias, vignetting, and exposure time normalization, along with rotating the images with a cubic convolution interpolation to place solar north at the top of the image. For COR1, this involved the extra step of individually background subtracting each polarization state before combining using a M\"ueller matrix to form polarized brightness images. For COR2, total brightness images were created and then studied as base difference images. Both COR1 and COR2 images were further enhanced using a wavelet technique \citep{byrn09}. 

\begin{figure*}[!ht]
\begin{center}
\includegraphics[height=\textwidth,angle=90]{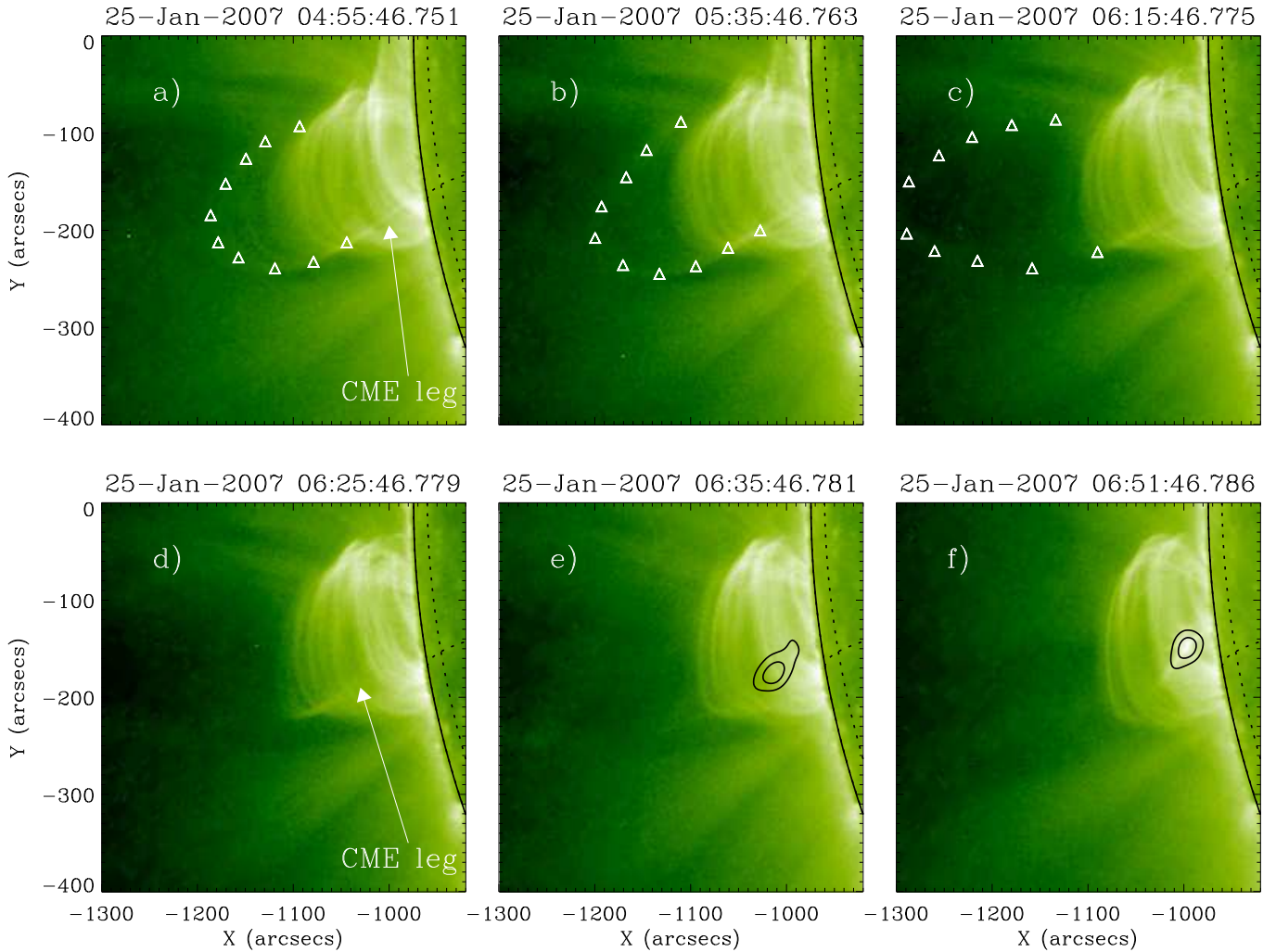}
\caption{Six select EUVI images in the 195~\AA~passband covering the time range 04:56--06:51~UT. Panels $a$--$c$ show the initial gradual rise of the CME front ($\it triangles$). A structure believed to be the southern leg of the CME is also noted. Overplotted on panels $e$ and $f$ are contours of the concurrent 5--10~keV emission observed by RHESSI, the plasmoid (source A) and looptop (source B), respectively. Note that the leg of the CME is no longer visible in these panels.}
\label{euvi_cme_front}
\end{center}
\end{figure*}

The CME front was first detected in the EUVI 195~\AA~passband at 04:56~UT at a height of $\sim$150~Mm above the eastern limb and gradually increased in height over the subsequent $\sim$1.5 hours (see Figures~\ref{euvi_cme_front}$a$--$c$). After 06:30~UT, when the CME became visible in COR1 images (as shown in Figure~\ref{euvi_cor1}), it began to expand more rapidly. At the same time a structure believed to be one leg of the CME (Figures~\ref{euvi_cme_front}$a$--$d$) was observed to sweep northwards to the site of the RHESSI 5--10~keV emission as noted in Figure~\ref{euvi_cme_front}e. This motion is interpreted as evidence for the predicted inflows associated with reconnection and will be discussed further in Section~\ref{inflows}. 

The maximum height of the CME front above the solar limb was measured in each frame to create a height-time profile. The assigned uncertainty in height of the CME front was taken to be five pixels, corresponding to uncertainties of 5, 50, and 200~Mm for EUVI, COR1 and COR2, respectively. From these, velocity and acceleration profiles along with their associated uncertainties were numerically derived using a three-point Lagrangian interpolation (see Figures~\ref{hsi_cme_ht}$a$--$c$) similar to that used by \cite{gall03}. This technique is not as sensitive to the uncertainties in the height-time measurements as a standard two-point numerical differentiation and can give an accurate representation of the acceleration profile. However, by smoothing the data in this way the magnitude of the values can only be taken as upper limits, at best. 

Figures~\ref{hsi_cme_ht}$a$--$c$ show that the CME front rose gradually for 1.5 hours with a mean velocity of $<$100~km~s$^{-1}$ before beginning to accelerate at $\sim$06:15~UT, when it was at a height of 250~Mm (1.35~R$_{\odot}$). The acceleration profile peaks some 20 minutes later when the CME was at a height of 400~Mm above the limb (1.57~R$_{\odot}$). Subsequently it continued to increase in height and velocity but at a decreasing rate. It obtained its maximum velocity of 1400~km~s$^{-1}$ at a height of 7000~Mm ($\sim$11~R$_{\odot}$) at $\sim$08:00~UT after which it began to decelerate. Figures~\ref{hsi_cme_ht}$d$ and \ref{hsi_cme_ht}$e$ show the height-time plot and lightcurves of the associated X-ray emission, respectively. It can be seen that the time of the observed downward motion of the plasmoid observed by RHESSI occurred during the acceleration phase of the CME. This lends further support to the idea that the CME front and the plasmoid were connected by a mutual current sheet and that the primary episode of reconnection both accelerated the CME and generated the magnetic tension in the field lines necessary to drive the plasmoid downwards. However, it is also possible that the CME acceleration was driven by external forces (e.g. kink instability in the flux rope) which led to filamentation of the current sheet and subsequent reconnection and plasmoid motion. 

\subsection{Reconnection Inflows}
\label{inflows}

During the initial gradual rise of the CME front, a linear structure believed to be the southern `leg' of the CME can be seen in panels $a$--$c$ of Figure~\ref{euvi_cme_front}. At 06:26~UT (Figure~\ref{euvi_cme_front}$d$) this structure was observed to sweep northwards towards the location of the RHESSI emission visible in the subsequent panel. It was no longer observed at its original location. Unfortunately, the northern leg was not visible, presumably obscured by the multitude of bright loops of the active region. 

To track the motion of this structure, a vertical one-pixel slice at Solar X = -1010$\arcsec$ was taken through a series wavelet enhanced EUVI images and stacked together in sequence to form a time series. The left-hand panel of Figure~\ref{euvi_time_slice} shows one such image taken at 06:51:47~UT with a solid vertical line denoting the pixel column used in the time series. The dotted line indicates the position of the CME leg some 3 hours earlier and the arrow denotes its inferred direction of motion. The solid contours mark the concurrent 5-10~keV emission observed by RHESSI (source B), which appeared elongated with a cusp to the southeast. A long narrow structure extending from the looptop to the southeast was also apparent in EUVI images. Such features are also often attributed to the reconnection process (e.g. \citealt{mcke99}). The emission associated with the plasmoid when it was first observed by RHESSI at 06:29~UT is also overlaid (source A; dashed contours) and is also located along the narrow structure in the EUVI image. 

The right-hand panel of Figure~\ref{euvi_time_slice} shows the time series of the one-pixel wide column through the series of wavelet-enhanced EUVI images. A feature was observed to propagate northwards from Solar Y $\approx-$195$\arcsec$ at 03:30~UT to Solar Y $\approx-$175$\arcsec$ at $\sim$06:50~UT, which was close to the site of the emission observed by RHESSI at that time. This time period also corresponds to the gradual rise phase of the CME front (noted in Figure~\ref{hsi_cme_ht}$a$). This feature is interpreted as evidence for the inflowing magnetic field lines associated with the slow reconnection prior to the main eruption. From this time series, an inflow velocity of 1.5~km~s$^{-1}$ was inferred, comparable to the 1.0--4.7~km~s$^{-1}$ value found by \cite{yoko01} using a similar method. Knowledge of the inflow velocity during a flare can provide information on the rate of reconnection and hence the energy release rate. The reconnection rate, $M_A$, is defined as the ratio of the inflow speed to the local Alfv\'{e}n speed. Taking a typical coronal Alfv\'{e}n speed of 1000~km~s$^{-1}$ the inflow velocity measured here would result in a value of $M_A$ = 0.001. This is also consistent with the lower end of the range of values for $M_A$ found by \citet{yoko01}. The brighter feature in the figure originating at solar Y $\approx$ -200$\arcsec$ and moving south is likely to be one of the active region loops being displaced as the CME erupts.

\section{DISCUSSION AND CONCLUSIONS}
\label{conc}

\begin{figure}[!t]
\begin{center}
\includegraphics[width=8.5cm]{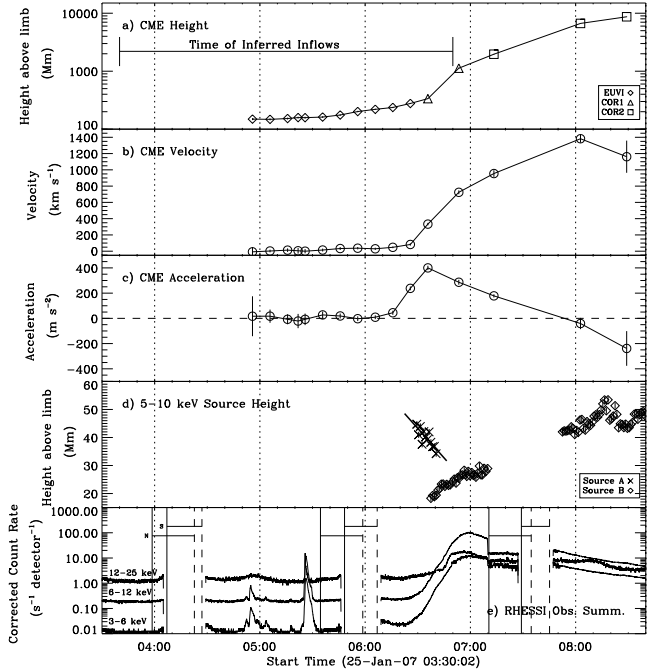}
\caption{Summary of the kinematics of both the CME observed with STEREO and the coronal X-ray sources observed with RHESSI. {\it a}) Height-time plot of the CME front from EUVI (diamonds), COR1 (triangles), and COR2 (squares). {\it b}) and {\it c}) The associated velocity and acceleration profiles, respectively. {\it d}) Height-time plot of the 5--10~keV sources as observed by RHESSI. The downward-moving coronal source is shown as crosses with error bars. The solid line denoted a least-squares fit to the data points and has been extended beyond the data points for clarity. The rising looptop source is represented by diamonds also with error bars. {\it e}) Observing summary profiles for RHESSI in the 3--6, 6--12 and 12--25~keV energy bands. Horizontal bars marked N and S denote RHESSI nighttimes and SAA passes, respectively.}
\label{hsi_cme_ht}
\end{center}
\end{figure}

\begin{figure}[!t]
\begin{center}
\includegraphics[height=8.5cm,angle=90]{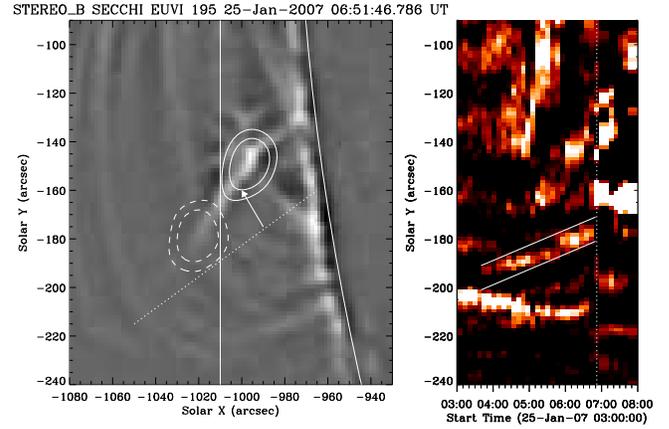}
\caption{{\it Left}: A wavelet enhanced EUVI image taken at 06:51:47~UT. The solid contours overlaid mark the position of the HXR looptop source at the time of the image. The dashed contour marks the position of the plasmoid at 06:29~UT. The dotted line shows the location of the CME leg at 04:46~UT and the arrow points in the direction of its motion. The solid vertical line denotes the pixel column used for the time-slice plot on the right. {\it Right}: A temporal evolution of a vertical slice through a series of EUVI images. The dotted line marks the time of the image in the left-hand panel. The structure believed to be the inflowing CME leg is identified between the two parallel lines.}
\label{euvi_time_slice}
\end{center}
\end{figure}

Rare observations are presented of a downward-propagating X-ray plasmoid appearing to merge with a looptop kernel during an eruptive event seen above the solar limb; the first case observed with RHESSI and perhaps only the second ever. Although the majority of above-the-looptop sources observed (in both white light and X-rays) tend to rise due to weaker magnetic field and decreasing density above the flare loops, in certain instances, conditions can be right for downward-moving plasmoids to form also. Enhanced HXR emission detected with RHESSI and radio emission observed by the Learmonth radio telescope suggest that this merging resulted in a secondary episode of particle acceleration (see Figure~\ref{hsi_ltc_ht_radio}). Images of the plasmoid formed over finer energy bins (as shown in Figure~\ref{hsi_ht_vs_en}) show that higher energy emission was observed at higher altitudes. This is consistent with the idea that the reconnection rate above the source was greater than that below, unlike rising plasmoids previously observed with RHESSI which show mean photon energy decreasing with height (e.g. \citealt{sui03}). Complementary observations from STEREO show that the plasmoid-looptop merging was concurrent with the period of the most rapid acceleration of the associated CME (Figures~\ref{hsi_cme_ht}$c$ and $d$). These observations are in agreement with a recent numerical simulation that predicts an increase in liberated energy during the merging of a plasmoid with a looptop source \citep{bart08a}. The formation of plasmoids is attributed to the tearing-mode instability during current sheet formation in the wake of an erupting CME \citep{furt63,lin00}.

\cite{bart07,bart08a} have shown theoretically that the deceleration of the plasmoid as it collides with the looptop source can lead to significant episodes of energy release. During this deceleration, antiparallel magnetic field lines begin to pile up between the two sources and a secondary current sheet is formed. This in turn leads to a secondary episode of magnetic reconnection that is driven by the magnetic tension of the field lines that govern the plasmoid motion. The authors also claim that the merging process triggers the excitation of large amplitude waves which can carry with them some of the stored magnetic energy. Although it is not possible to detect any acceleration or deceleration from the RHESSI images presented, a mean downward velocity of 12~km~s$^{-1}$ was calculated. This value is commensurate with the previous observation of \cite{kolo07}, who measured 16~km~s$^{-1}$ during a similar event observed with Yohkoh. However, both these observed values are considerably lower than the value predicted by \cite{bart08a} of $\sim$40\% of the local Alfv\'{e}n speed (i.e. $\sim$400~km~s$^{-1}$). Similar values of $\sim$200~km~s$^{-1}$ were predicted by \cite{rile07} for downward-moving white-light plasmoids. The low velocity measured here may be attributed to the low value of the reconnection rate as estimated from the inflows observed with EUVI (assuming that these field lines converged above the plasmoid). The value of $M_A \approx$ 0.001 is an order of magnitude lower than that used in the numerical simulation. As the amount of tension exerted on the plasmoid is sensitive to the net reconnection rate, this would result in a lower tension force and therefore lower downward velocity. This in turn may also affect the amount of energy liberated in the subsequent collision with the looptop. It is possible that the model of \cite{bart08a} may overestimate the velocity (and subsequent dissipated energy) given that the simulation is two-dimensional and does not take into account 3D structures, such as a twisted flux rope. Similarly the plasmoid detected with RHESSI is observed for more than 10 minutes before merging with the looptop source, whereas the simulations which yielded higher velocities predict that the source should exist for only $\sim$1 minute before merging.  While the simulations of \cite{bart08a} predict a rebrightening of the loop footpoints in HXRs and/or chromospheric emission, both the analysis presented here and that of \citealt{kolo07} show a distinct increase in coronal emission. A recent analysis of Miklenic et al. (2010; submitted) appears to refute the idea that plasmoid-looptop interactions could be responsible for chromospheric rebrightenings. These observations provide further evidence that the particle acceleration process occurs in the corona rather than at the footpoints as recently suggested by \cite{flet08}, although acceleration at the footpoints as recently suggested by \cite{brow09} cannot be ruled out. 

While plasmoid-looptop interactions are rarely observed, it is possible that they occur more often but are difficult to observe due to the brighter emission from the flare itself and RHESSI's limited dynamic range. A newly developed technique of deep integrations using RHESSI visibility-based X-ray imaging \citep{sain09} may help to identify faint X-ray sources in the corona during eruptive limb events. By comparing other similar events it may be possible to determine how great an effect the CME acceleration (and magnetic reconnection rate, if possible) has upon the resulting HXR and radio flux. It would therefore be useful to find events observed at a time when RHESSI's detector calibration was better known in order to perform a more rigorous spectral analysis which was not possible for this event due to poorly known calibration. This could reveal more detailed information on the energetics of the resulting accelerated particles.

\acknowledgements
ROM would like to thank Gordon Holman and Jack Ireland for their very helpful and insightful discussions, and Kim Tolbert for modifications made to the RHESSI software. We also thank the anonymous referee for their constructive comments which improved the quality of this paper. RTJMA is funded by a Marie Curie Intra European Fellowship. CAY acknowledges support from NASA Heliophysics Guest Investigator grand NNG08EL33C.


\bibliographystyle{apj}
\bibliography{ms_preprint}

\end{document}